\pdfoutput=1

\documentclass[11pt, letterpaper]{article}
\usepackage[margin={2cm,2cm}]{geometry} 

\usepackage{sectsty}
\usepackage{abstract}
\usepackage{color}
\usepackage{graphicx}
\usepackage{multicol}
\usepackage{amsmath,amssymb,amstext,amsfonts}
\usepackage[usenames,svgnames]{xcolor}
\usepackage{float}
\usepackage{subfig}
\usepackage{caption}
\usepackage{framed}
\usepackage{authblk}
\usepackage{titlesec}
\usepackage{setspace}
\usepackage{times,fancyhdr}
\usepackage[normalem]{ulem} 
\usepackage{enumerate} 
\usepackage{mathrsfs} 
\usepackage{bigints}
\usepackage{bm}
\usepackage{enumitem} 
\usepackage[hypertexnames=false]{hyperref}

\titlespacing*{\section}{0pt}{3ex plus 2ex}{1ex} 
\sectionfont{\fontsize{11}{11}\selectfont} 
\subsectionfont{\fontsize{10}{10}\selectfont} 

\hypersetup{
    colorlinks=true,
    urlcolor=SteelBlue,
    linkcolor=red,
    citecolor=blue,
}

\newcommand*{\Scale}[2][4]{\scalebox{#1}{$#2$}} 
\newcommand{\romansubs}{\renewcommand{\theequation}{\theparentequation \roman{equation}}} 
\newcommand{\tinymath}[1]{\mbox{\tiny{$#1$}}}
\newcommand{\tinyrmsub}[1]{\mbox{{\tiny{#1}}}}
\newcommand{\alphaone}{\alpha\tinymath{1}}
\newcommand{\alphatwo}{\alpha\tinymath{2}}

\newcommand{\PRLsep}{\noindent\makebox[\linewidth]{\resizebox{0.750\linewidth}{1pt}{$\blacklozenge$}}\bigskip}
\newcommand{\dd}{\mathrm{d}}

\begin{document}
\pagestyle{fancy}
\fancyhead{} 
\fancyhead[OR]{\thepage}
\fancyhead[OC]{{\small{
   \textsf{Regular solutions in $f(T)$-Yang-Mills theory}}}}
\fancyfoot{} 
\renewcommand\headrulewidth{0.5pt}
\addtolength{\headheight}{2pt} 
\global\long\def\tdud#1#2#3#4#5{#1_{#2}{}^{#3}{}_{#4}{}^{#5}}
\global\long\def\tudu#1#2#3#4#5{#1^{#2}{}_{#3}{}^{#4}{}_{#5}}

\twocolumn

\title{\vspace{-2cm}\hspace{-0.0cm}\rule{\linewidth}{0.2mm}\\
\bf{\Large{\textsf{Regular solutions in $\bm{f(T)}$-Yang-Mills theory}}}}


\author[1,2]{\small{Andrew DeBenedictis}\thanks{\href{mailto:adebened@sfu.ca}{adebened@sfu.ca}}}%

\author[3]{\small{Sa{\v s}a Iliji{\'c}}\thanks{
   \href{mailto:sasa.ilijic@fer.hr}{sasa.ilijic@fer.hr}}}%


\affil[1]{\footnotesize{\it{The Pacific Institute
   for the Mathematical Sciences}} \protect\\
   \footnotesize{and}
}

\affil[2]{\footnotesize{\it{Department of Physics, Simon Fraser University}}\\
   \footnotesize{\it{8888 University Drive, Burnaby, BC, V5A 1S6, Canada}}
}

\affil[3]{
   \footnotesize{\it{Department of Applied Physics,
   Faculty of Electrical Engineering and Computing, University of Zagreb}}\\
   \footnotesize{\it{HR-10000 Zagreb, Unska 3, Croatia}}
}

\date{\vspace{-0.8cm}(\footnotesize{July 9, 2018})} 
\twocolumn[ 
  \begin{@twocolumnfalse}  
  \begin{changemargin}{1.75cm}{1.75cm} 
\maketitle
\end{changemargin}
\vspace{-1.0cm}
\begin{changemargin}{1.5cm}{1.5cm} 
\begin{abstract}
{\noindent\small{We consider extended covariant teleparallel $(f(T))$ gravity whose action is analytic in the torsion scalar and which is sourced by an $su(2)$ valued Yang-Mills field. Specifically, we search for regular solutions to the coupled $f(T)$-Yang-Mills system. For $f(T)=T$, not surprisingly, the Bartnik-McKinnon solitons of Einstein Yang-Mills theory are recovered. However, interesting effects are discovered with the addition of terms in the action which are nonlinear in the torsion scalar, which are specifically studied up to cubic order. With the addition of the nonlinear terms the number of regular solutions becomes finite. As well, beyond critical values of the coupling constants it is found that there exist \emph{no} regular solutions. These behaviors are asymmetric with respect to the sign of the nonlinear coupling constants and the elimination of regular solutions turns out to be extremely sensitive to the presence of the cubic coupling. It may be possible, therefore, that with sufficiently high powers of torsion in the action, there may be no regular Yang-Mills static solutions.}}
\end{abstract}
\noindent{\footnotesize PACS(2010): 04.50.Kd \;\; 12.10.-g \;\; 12.15.-y}
\rule{\linewidth}{0.2mm}
\end{changemargin}
\end{@twocolumnfalse} 
]
\saythanks 
\vspace{0.5cm}
{\setstretch{0.9} 
\section{Introduction}
It is well known that general relativity, with its curvature description of the gravitational interaction, is a highly successful theory of gravity. Interestingly there is an equivalent theory which is purely torsion based, and therefore possesses no curvature. This other theory has become known as the teleparallel equivalent of general relativity (TEGR). This torsion theory of gravity, along with its extensions, known as extended teleparallel gravity or $f(T)$ gravity has gained much interest in the past few decades within the realm of modified gravity theories. Arguably the greatest amount of work has been performed in cosmology \cite{ref:sharifandrani} - \cite{ref:cosconstr} and it has been shown that $f(T)$ gravity may be able to naturally yield dark matter and dark energy effects \cite{ref:dmstart}-\cite{ref:dmend}. As well, $f(T)$ gravity has been applied to the study of relativistic stellar structure \cite{ref:tamgood} - \cite{ref:sasamarco}  and, to a lesser extent, black holes \cite{ref:bh1} - \cite{ref:bhend}. In relevance to electromagnetism, teleparallel gravity has been used in a number of interesting studies \cite{ref:elecbh}, \cite{ref:sasaelec}, \cite{ref:chargegrava}. 

To date, we are unaware of any studies dealing with non-Abelian fields coupled to teleparallel gravity. In the realm of general relativity the Einstein Yang-Mills (EYM) system possesses some fascinating features \cite{ref:richnesseym}. There are, for example, the famous discrete regular solutions of Bartnik and McKinnon \cite{ref:BM} where it is found that an infinite family of regular magnetic solutions exist to the static spherically-symmetric EYM system but only for discrete values of the ``tuning parameter'' \cite{ref:infinite1}, \cite{ref:infinite2}, chosen by Bartnik and McKinnon to be the second derivative of the Yang-Mills potential. This sort of behavior was found earlier in the Einstein-Maxwell-Klein-Gordon system by Das and Coffman \cite{ref:dascoff}. Since then work has been extended to include Higgs fields \cite{ref:galthiggs}-\cite{ref:jiahiggs}, and symmetries relaxed to the more general axial symmetry as well as non asymptotically flat solutions \cite{ref:pertaxial} - \cite{ref:breitlambda}. Some studies have also been done regarding Einstein Yang-Mills black holes \cite{ref:volgal} - \cite{ref:volkovreview}. Some work has also been done within certain theories of quantum gravity \cite{ref:mason} as well as in curvature based modified gravity \cite{ref:mazharimousavi}.

We wish here to bring the study of non-Abelian Yang-Mills fields to the realm of teleparallel gravity. Specifically we consider an $f(T)$ action which may either be considered exact, or a representation of low-order terms in a more general action analytic in torsion. To this an $SU(2)$ Yang-Mills field is minimally coupled and regular static solutions to the resulting system of equations are sought. By regular we mean that there is no singularity in the Yang-Mills electric and magnetic fields, and no singularity in the spacetime. We also do not consider cases with horizons. In TEGR the regular solitons of Bartnik-McKinnon are recovered, which is to be expected given its equivalence to general relativity. However, some interesting differences manifest when the theory deviates from TEGR: The infinite family of regular solutions becomes finite and, beyond critical values of the nonlinear coupling constants there are no regular solutions. This feature seems to be particularly sensitive to the presence of higher order terms in the action.

\subsection[Covariant $f(T)$ gravity in brief]{Covariant $\bm{f(T)}$ gravity in brief}
The extended teleparallel action is given by \footnote{The index structure in this paper is such that unadorned Greek letters are utilized for spacetime indices, hatted Greek letters for orthonormal indices, and Latin letters for $SU(2)$ group indices.} 
\begin{equation}
S=\int \left(\frac{1}{16\pi}\, f(T)  +\mathcal{L}\tinyrmsub{matter}\right){\mbox{det}(h^{\hat{\alpha}}_{\;\;\mu})}\,d^{4}x\,, \label{eq:tegraction}
\end{equation}
where $h^{\hat{\alpha}}_{\;\;\mu}$ represents the tetrad, which must meet the metric compatibility condition
\begin{equation}
h^{\hat{\alpha}}_{\;\;\mu}h_{\hat{\alpha}\nu}=g_{\mu\nu}\,. \label{eq:metcomp}
\end{equation}
The quantity $f(T)$ is some function of the torsion scalar, which is constructed out of the torsion tensor, $T^{\alpha}_{\;\;\beta\gamma}$. This, in turn, is defined from the commutator of the flat Weitzenb\"{o}ck connection $\Gamma^{\sigma}_{\;\;\beta\gamma}$, supplemented with spin connection, $\omega^{\hat{\alpha}}_{\;\;\hat{\beta}\sigma}$ via
{\allowdisplaybreaks\begin{align}
T^{\hat{\alpha}}_{\;\;\mu\nu}=&h^{\hat{\alpha}}_{\;\;\sigma}\left(\Gamma^{\sigma}_{\;\;\nu\mu}-\Gamma^{\sigma}_{\;\;\mu \nu}\right) = \partial_{\mu}h^{\hat{\alpha}}_{\;\;\nu} - \partial_{\nu}h^{\hat{\alpha}}_{\;\;\mu} \nonumber \\
& + \omega^{\hat{\alpha}}_{\;\;\hat{\beta}\mu}h^{\hat{\beta}}_{\;\;\nu} - \omega^{\hat{\alpha}}_{\;\;\hat{\beta}\nu}h^{\hat{\beta}}_{\;\;\mu}\,.\label{eq:gentor}
\end{align}}
Explicitly, the torsion scalar itself is constructed as:
\begin{equation} 
    T :=
    \frac14  T_{\alpha\beta\gamma}  T^{\alpha\beta\gamma}
     + \frac12  T_{\alpha\beta\gamma}  T^{\gamma\beta\alpha}
     - T_{\alpha\beta}^{\;\;\;\alpha}T^{\gamma\beta}_{\;\;\;\;\gamma}\,, \label{eq:genT}
\end{equation}

Before continuing, the perhaps unfamiliar presence and role of the spin connection above requires some discussion. Without the spin connection one has the ``traditional'' form of teleparallel gravity. As long as one is limited to the action linear in $T$ there is no harm in ignoring the spin connection and working within the traditional form of the theory. However, it was soon realized that if one wishes to extend the Lagrangian to include terms nonlinear in $T$, forming what is known as extended teleparallel gravity, one runs into the issue of violation of local Lorentz covariance \cite{ref:goodbadtets} - \cite{ref:davood}. That is, the equations of motion, along with physical observables, cannot generally be formulated in a Lorentz covariant manner. This then requires one to search for a particular tetrad, which obeys the condition (\ref{eq:metcomp}) but which is also compatible with a vanishing spin connection. These are known as ``good'' tetrads \cite{ref:goodbadtets}. This presents a serious drawback to the traditional theory in that metric compatibility (via (\ref{eq:metcomp})) is an insufficient criterion for the tetrad and one runs into the possibility of having a metric compatible tetrad which does not yield Lorentz covariant equations of motion. It has been generally realized that to restore full covariance to the theory, the spin connection cannot be ignored and must play a role  \cite{ref:earlycov}, \cite{ref:earlycov2}, \cite{ref:davood}. Some interesting progress has been made on this front in the last few years \cite{ref:KandS}-\cite{ref:2018cov}. 

As we will be working with extended teleparallel Lagrangians, the spin connection will be utilized so that we may use a simple tetrad, and therefore the requirement of the concern for whether it is a ``good'' or ``bad'' tetrad is eliminated. The method introduced in \cite{ref:KandS} to restore covariance is robust in static spherical symmetry in curvature coordinates, and hence we will utilize the scheme in \cite{ref:KandS} in order to calculate the spin connection. In brief, if a line-element is used of the form
\begin{equation}
\Scale[0.90]{ds^{2}= -A^{2}(r)\, dt^{2} + B^{2}(r)\, dr^{2} + r^{2}\, d\theta^{2} +r^{2}\sin^{2}\theta\, d\phi^{2}}\,, \label{eq:lineelement}
\end{equation}
with $r$ the usual radial coordinate, then one can calculate the appropriate spin connection (see \cite{ref:KandS}, \cite{ref:ourvac}, \cite{ref:sasamarco} for details) as:
\begin{align}
\omega^{\hat{r}\hat{\theta}}{}_\theta = - \omega^{\hat{\theta}\hat{r}}{}_\theta = 1, \nonumber \\
\omega^{\hat{r}\hat{\phi}}{}_\phi = - \omega^{\hat{\phi}\hat{r}}{}_\phi = \sin\theta, \nonumber \\
\omega^{\hat{\theta}\hat{\phi}}{}_\phi = - \omega^{\hat{\phi}\hat{\theta}}{}_\phi = \cos\theta
\,. \label{eq:ourspincon}
\end{align}
With the above spin connection one may choose any metric compatible tetrad for static spherical symmetry without the worry of losing local Lorentz covariance. We therefore choose the following diagonal tetrad:
\begin{equation}
\left[h^{\hat{\alpha}}_{\;\;\mu}\right]=\left( \begin{array}{cccc}
A(r) & 0 & 0 & 0\\
0 & B(r) & 0 & 0 \\
0 & 0 & r & 0 \\
0 & 0 & 0 & r\sin\theta  \end{array} \right)\,. \label{eq:diagtet}
\end{equation}

Using the spin connection (\ref{eq:ourspincon}), the torsion scalar is
\begin{equation}
T=\frac{2(B-1)(A-AB +2rA')}{r^{2}AB^{2}}\,. \label{eq:mettors}
\end{equation}
Again it should be stressed that the torsion scalar (\ref{eq:mettors}) in the case where one properly calculates the spin connection, $\omega^{\hat{\mu}\hat{\nu}}_{\;\;\;\sigma}$, is the proper Lorentz invariant torsion scalar for static spherical symmetry in the coordinate system (\ref{eq:lineelement}). If one ignores the spin connection, then one must be sure to use a tetrad which yields zero spin connection, in which case one would obtain the same torsion scalar as (\ref{eq:mettors}). That is, the tetrad must be properly ``parallelized''. If one uses an improperly rotated tetrad one will find that the spin connection, calculated via the method of \cite{ref:KandS}, is not zero, and cannot be ignored. If it is ignored one will get a different torsion scalar than (\ref{eq:mettors}) which will depend on Lorentz frame. Examples of properly rotated tetrads may be found in \cite{ref:tamgood}, \cite{ref:KandS}, \cite{ref:tampres}.

In the following section we will use tetrad (\ref{eq:diagtet}) in order to create the gravitational sector of the equations of motion. The Yang-Mills sector will also be constructed and, with appropriate Yang-Mills ansatz, we will solve the equations for an $f(T)$ consisting of several nonlinear terms which comprise the lowest order terms in an analytic (in $T$) Lagrangian density. This will be followed with an in depth analysis of the system of equations, subject to conditions of regularity, and comments on several interesting results which are found. We do not study the stability of these solutions, due to the obstruction of incorporating time dependent perturbations in extended teleparallel gravity. This difficulty arises from the fact that at present there is no straightforward method of obtaining a spin connection which allows for proper time-dependent equations in the spherical coordinate chart within extended teleparallel gravity, although some progress is being made \cite{ref:KandS}-\cite{ref:2018cov}. Finally we will make some concluding remarks.

\section[The $f(T)$-Yang-Mills system]{The $\bm{f(T)}$-Yang-Mills system}\label{sec:fym}
We wish to study here $SU(2)$ Yang-Mills fields coupled minimally to $f(T)$ gravity. Specifically, the action taken is
\begin{equation}
\Scale[0.99]{S=\frac{1}{16\pi}\bigintss \left(f(T)+F_{i\mu\nu}F^{i\mu\nu}\right){\mbox{det}(h^{\hat{\alpha}}_{\;\;\mu})}\,d^{4}x}\,, \label{eq:nonlinact}
\end{equation}
with
\begin{equation}
f(T)= T + \frac{\alphaone}{2}T^{2} + \frac{\alphatwo}{6}T^{3}\,. \label{eq:ourfoft}
\end{equation}
The non-Abelian field strength tensor,  $F^{i}_{\;\;\mu\nu}$, is given in terms of the $SU(2)$ potential, $W^{i}_{\;\;\mu}$, as:
\begin{equation}
F^{i}_{\;\;\mu\nu}=\partial_{\mu}W^{i}_{\;\;\nu}-\partial_{\nu}W^{i}_{\;\;\mu}-e \epsilon^{i}_{\;jk} W^{j}_{\;\;\mu} W^{k}_{\;\;\nu}\,. \label{eq:fst}
\end{equation}
Here $e$ is the self coupling charge of the non-Abelian field.

Before proceeding we should comment on a possible confusion with (\ref{eq:fst}). Strictly speaking, the partial derivatives in (\ref{eq:fst}) should be spacetime covariant derivatives, which in turn introduce the connection into the field strength tensor. If the connection is not symmetric, a torsion term will appear in the resulting field strength tensor. However, in teleparallel gravity it is known that gauge invariance of the field-strength and its equations of motion require the use of the Fock-Ivanenko derivative \cite{ref:torem}, \cite{ref:booktor}, which in turn yields a field strength tensor equivalent to the one calculated with the \emph{Christoffel connection}. Therefore the field-strength tensor (\ref{eq:fst}) is actually identical to the one in Einstein Yang-Mills theory.

For the field potential we adopt a slight variation of Witten's spherically symmetric $SU(2)$ ansatz \cite{ref:witten}, \cite{ref:zhou}, which after gauge fixing may be written as:
\begin{align}
\Scale[0.9]{W=}&\Scale[0.9]{W^{i}_{\;\;\mu}\,\tau_{i}\,dx^{\mu}} \nonumber \\
=& \Scale[0.9]{V(r)\,\mathrm{d}t\,\tau_{r}+W(r)\left[\mathrm{d}\theta\,\tau_{\theta}
- \sin \theta\, \mathrm{d}\phi\,\tau_{\phi}\right] +\frac{1}{e}\cos\theta\, \mathrm{d}\phi\,\tau_{r}\,.} \nonumber
\end{align}
The $\tau_{j}$ generators are proportional to the (spherical projection) Pauli matrices obeying the algebra
\begin{equation}
\left[\tau_{j},\,\tau_{k}\right]=i \epsilon_{jk}^{\;\;\;\,\ell}\,\tau_{\ell}\,.
\end{equation}

From the action (\ref{eq:nonlinact}) one may calculate the gravitational equations of motion via variation of the tetrad as
\begin{align}
& h^{-1} h^{\hat{\alpha}}{}_{\mu} \partial_{\sigma}
     \left( h \frac{\dd f(T)}{\dd T} S_{\hat{\alpha}}{}^{\nu\sigma} \right) \nonumber \\
& - \frac{\dd f(T)}{\dd T} T_{\alpha\beta\mu} S^{\alpha\beta\nu}
+ \frac12 f(T) \delta_{\mu}{}^{\nu} \nonumber \\
& + \frac{\dd f(T)}{\dd T}
          S_{\hat{\alpha}}{}^{\sigma\nu}
          h^{\hat{\beta}}{}_{\mu}
          \omega^{\hat{\alpha}}{}_{\hat{\beta}\sigma}
 = 8\pi \mathcal{T}_{\mu}{}^{\nu}\,, \label{eq:graveoms}
\end{align}
with $h$ the determinant of the tetrad and $\mathcal{T}_{\mu}{}^{\nu}$ the usual symmetric stress-energy tensor
\begin{equation}
\mathcal{T}_{\mu}{}^{\nu} = \frac{1}{4\pi} \left(
F_{i\mu\alpha} F^{i\nu\alpha}
- \frac{1}{4} \delta_{\mu}{}^{\nu} F_{i\alpha\beta} F^{i\alpha\beta} 
\right)\,. \label{eq:setensor}
\end{equation}
In (\ref{eq:graveoms}) the quantity sometimes known as the superpotential is present: 
\begin{equation} 
    S_{\alpha\beta\gamma} :=  K_{\beta\gamma\alpha}
      + g_{\alpha\beta} \,  T_{\sigma\gamma}{}^{\sigma}
      - g_{\alpha\gamma} \,  T_{\sigma\beta}{}^{\sigma}\,, \label{eq:superpot}
\end{equation}
where $K_{\beta\gamma\alpha}$ is the contorsion tensor:
\begin{equation} \label{eq:ktensor}
    K_{\alpha\beta\gamma} :=
     \frac12 \left(  T_{\alpha\gamma\beta}
     +  T_{\beta\alpha\gamma} +  T_{\gamma\alpha\beta} \right)\,.
\end{equation}

As well, variation with respect to the Yang-Mills potential yields the Yang-Mills equations of motion\footnote{In teleparallel gravity the stress-energy tensor turns out to be conserved with respect to the Christoffel connection. It can be shown that this conservation law and the Yang-Mills equations of motion, under mild assumptions, imply each other.}
\begin{equation}
D_\nu F_i{}^{\nu\beta} = 0\,. \label{eq:ymeoma}
\end{equation}
where the operator $D$ is the gauge covariant derivative
\begin{equation}
D_\mu F_{i\nu\lambda} := F_{i\nu\lambda;\mu}
                      - e \; \epsilon_{bci} W^{b}{}_{\mu} F^{c}{}_{\nu\lambda} \,.\label{eq:ymeomb}
\end{equation}
Here the semi-colon denotes a covariant derivative with respect to the \emph{Christoffel connection} (again due to the Fock-Ivanenko derivative).

Explicitly for the case studied here, equations (\ref{eq:ymeoma}) are
\begin{subequations}
\romansubs
{\allowdisplaybreaks\begin{align}
\Scale[0.98]{W^{\prime\prime}=} & \Scale[0.98]{-\frac{A^{\prime} W^{\prime}}{A}+B^2 W \left(\frac{e^2 W^2-1}{r^2}-\frac{e^2
   V^2}{A^2}\right)}  \nonumber \\
&\Scale[0.98]{+\frac{B^{\prime} W^{\prime}}{B}} , \label{eq:ymexplicita} \\[0.2cm]
\Scale[0.98]{V^{\prime\prime}=}& \Scale[0.95]{\left(\frac{A^{\prime}}{A}-\frac{2}{r}\right)
   V^{\prime}+\frac{B^{\prime} V^{\prime}}{B}+\frac{2 e^2 B^2 V W^2}{r^2}}. \label{eq:ymexplicitb}
\end{align}}
\end{subequations}
The gravitational equations of motion are rather complicated and are therefore presented in the Appendix. From (\ref{eq:setensor}) the quantities $\rho$, $p$ and $q$ may be acquired, which are the energy density, radial, and transverse pressure respectively of the Yang-Mills field:
\begin{subequations}
\romansubs
{\allowdisplaybreaks\begin{align}
\rho =& \Scale[0.95]{\frac{1}{8 \pi  e^2 r^4 A^2 B^2}
\Big[A^2 \left(B^2 \left(e^2
   W^2-1\right)^2+2 e^2 r^2 W'^2\right) } \nonumber \\
 &\Scale[0.95]{ +2 e^4 r^2 B^2 V^2
   W^2+e^2 r^4 V'^2\Big]} \\
p =& \Scale[0.95]{ -\frac{1}{8 \pi  e^2 r^4 A^2 B^2} \Big[ A^2
\left(B^2 \left(e^2
   W^2-1\right)^2-2 e^2 r^2 W'^2\right) } \nonumber \\
&\Scale[0.95]{ -2 e^4 r^2 B^2 V^2
   W^2+e^2 r^4 V'^2\Big]}\\
q = & \Scale[0.95]{\frac{1}{8 \pi e^2 r^4 A^2 B^2}\Big[A^2 B^2 \left(e^2
   W^2-1\right)^2}\Scale[0.95]{+e^2 r^4 V'^2\Big]}\,.
\end{align}}
\end{subequations}

At this stage one has all the quantities required in order to solve the equations, save for the boundary conditions. We next concentrate on obtaining these conditions as well as the solutions.

\section{Solutions}\label{sec:sols}
In order to obtain computational solutions to the previous equations, we require a system of independent equations from which the highest derivative of the quantities to be evolved can be extracted. Below the assumption is made that all quantities are Laurent expandable.

\subsection{Initial conditions and asymptotics}
At this stage we require initial conditions for the evolutions. The evolutions cannot start exactly at $r=0$ due to the non-essential singularity present there. However, we can start the evolution slightly away from $r=0$. This then requires that we have initial values at this point. Since the starting point is near $r=0$, accurate values for these initial conditions may be obtained from an expansion to some appropriate order in $r$ about $r=0$. Then all initial data simply requires knowledge of the various functions at $r=0$, and we have sufficient criteria to restrict these values. There is a rather large number of conditions required (both at $r=0$ and, as well, required at ``infinity'' in order to have a physically reasonable solution). Due to the number of different requirements, we refer the reader to table 1 for an overview. A number of these conditions are actually redundant, but are included to ensure that all reasonable criteria are considered. The various criteria and their motivation are explained as follows:

\begin{table*}[!t!]
\begin{center}
\begin{tabular*}{\textwidth}{|p{2.6cm}||p{1.20cm}|p{1.20cm}|p{1.20cm}|p{1.20cm}|p{1.20cm}|p{1.20cm}|p{1.20cm}| p{1.20cm}|p{1.20cm}|p{1.20cm}|p{1.20cm}|} 
 \hline
 \multicolumn{12}{|c|}{Table 1: Restrictions on various quantities} \\
 \hline
 {\small{Condition}} & $A'$ & $A''$ & $B$ & $B'$ & $B''$ & $W$ & $W'$ & $W''$ & $V$ & $V'$ & $V''$ \\
 \hline
 \small{All equations solved near $r=0$}  & 0    &$\frac{V'^2}{A} +AW''^{2}$& 1 & 0 &$\frac{V'^2}{A}+W''^{2}$ & $\pm {1}/{e}$ & 0 &  & 0 & & 0 \\
  \hline
\small{Finite $\mathcal{T}^{\hat{\mu}\hat{\nu}}(0)$}  &     & &  &  & & $\pm{1}/{e}$ & 0 &  & 0 & &  \\
  \hline
\small{${}^{*}\;\Scale[1.00]{\mathcal{T}^{\hat{\mu}\hat{\nu}}(r\neq 0)=0}$}  &     & &  &  & & $\pm {1}/{e}$ &  &  & 0 & &  \\
  \hline
\small{Finite $\Scale[1.00]{F^{i}_{\;\;\hat{\mu}\hat{\nu}}(0)}$}  &  & &  &  & & $\pm {1}/{e}$ & 0 &  & 0 & &  \\
  \hline
\small{${}^{*}\;\Scale[0.95]{F^{i}_{\;\;\hat{\mu}\hat{\nu}}(r \neq 0)=0}$}  &     & &  &  & & $\pm {1}/{e}$ & 0 &  & 0 & 0 &  \\
  \hline
\small{Finite $R^{\hat{\alpha}}_{\;\hat{\beta}\hat{\gamma}\hat{\delta}}(0)$}  &  0  & & 1 & 0 & &  &  &  & & &  \\
  \hline
\small{${}^{*}\;\Scale[0.92]{R^{\hat{\alpha}}_{\;\hat{\beta}\hat{\gamma}\hat{\delta}}(r \neq 0)=0}$}  &  0  & 0 & 1 & 0 & &  &$ie\frac{VW}{A}$  &  & & ${\Scale[0.50]{\frac{iA}{er^{2}}(e^{2}W^{2}-1)}}$ &  \\
  \hline
\small{Finite $T^{\hat{\alpha}}_{\;\hat{\beta}\hat{\gamma}}(0)$}  &   & & 1 &  & &  &  &  & & &  \\
  \hline
\small{${}^{*}\;\Scale[1.00]{T^{\hat{\alpha}}_{\;\hat{\beta}\hat{\gamma}}(r)=0}$}  &  0  &  & 1 &  &  &  &  & 
$\Scale[0.45]{W\left(\frac{e^{2}W^{2}-1}{r^{2}}\right.} -\Scale[0.45]{\left.\frac{e^{2}V^{2}}{A^{2}}\right)+B'W'}$ & & 0 & \\
  \hline
\multicolumn{12}{l}{{\small{${}^{*}$These conditions are sufficient (but not always necessary) to make the quantity vanish as $r \rightarrow \infty$.}}} \\
\hline
\end{tabular*}
\end{center}
\end{table*}


\begin{itemize}[leftmargin=*]
  \item {\bf{All equations solved near}} $\bm{r=0}:$ This is a requirement that near $r=0$ (close to where the evolution starts) the left-hand side of all equations of motion equal, to sufficient order in $r$, the right-hand side. This is done by considering equality of the l.h.s. to the r.h.s. up to sufficiently high order in the expansion of the equations of motion. By setting the expansion coefficients of the l.h.s. at fixed order equal to the corresponding expansion coefficient on the r.h.s. one may determine what the relationship between various quantities appearing in the expansion coefficients are. Since expansion coefficients of any quantity near $r=0$ are constructed from $r=0$ data, we can use $r=0$ data to form these coefficients. The $r=0$ data required is determined from some of the criteria listed below.
  \item {\bf{Finite}} $\bm{\mathcal{T}^{\hat{\mu}\hat{\nu}}(0)}:$ This is a requirement that the physical observables $\rho$, $p$ and $q$ do not become infinite at the origin. These conditions are satisfied by demanding that the coefficients of negative powers of $r$ in the expansions of $\rho$, $p$ and $q$ vanish. This is both a physical requirement as well as a way to determine initial conditions.
  \item $\bm{\mathcal{T}^{\mu\nu}(r\neq 0)=0}:$ This condition can be used as $r \rightarrow \infty$ and demands that the stress-energy of the Yang-Mills field vanishes as one approaches infinity. This is a physical condition and not related to regularity nor directly related to initial conditions.
\item {\bf{Finite}} $\bm{\mathcal{F}^{i}_{\;\hat{\mu}\hat{\nu}}(0)}:$ This is a requirement that all electric and magnetic fields (in principle observables) be finite at the origin. We satisfy these conditions by demanding that the coefficients of negative powers of $r$ in the expansions of the components of $F^{i}_{\;\;\hat{\mu}\hat{\nu}}$ vanish. This is both a physical requirement as well as a way to determine initial conditions. This condition also implies the earlier condition of regularity of the stress-energy tensor.
 \item $\bm{\mathcal{F}^{i}_{\;\hat{\mu}\hat{\nu}}(r \neq 0)=0}:$ This condition can be used as $r \rightarrow \infty$ and demands that the electric and magnetic fields vanish as one approaches infinity. This is a physical condition and not related to regularity nor directly related to initial conditions.
\item  {\bf{Finite Christoffel}} {$\bm{R^{\hat{\alpha}}_{\;\hat{\beta}\hat{\gamma}\hat{\delta}}(0):}$} It may seem peculiar that in a theory with no curvature one makes demands on the Riemann-Christoffel tensor. However, in $f(T)$ gravity it can be shown that the force equation on test particles is identical to the geodesic equation calculated with the Christoffel connection \cite{ref:booktor}. Therefore, particles subject to the force of gravity in $f(T)$ theory are subject to the Riemann-Christoffel geodesic deviation equation. This therefore is a condition of demanding finite tidal forces even in $f(T)$ gravity.
\item {\bf{Christoffel}} {$\bm{R^{\hat{\alpha}}_{\;\hat{\beta}\hat{\gamma}\hat{\delta}}(r\neq 0) =0:}$} For the analogous reason as in the previous item we consider the Riemann-Christoffel tensor as $r\rightarrow \infty$. Here, this is a requirement demanding tidal forces vanish there.
\item  {\bf{Finite}} {$\bm{T^{\hat{\alpha}}_{\;\hat{\beta}\hat{\gamma}}(0):}$} In $f(T)$ gravity the torsion tensor (via the contorsion) plays the role of the gravitational force \cite{ref:booktor}. We therefore demand that the gravitational force not be infinite.
\item {$\bm{T^{\hat{\alpha}}_{\;\hat{\beta}\hat{\gamma}}(r \neq 0) =0:}$} This condition can be used as $r \rightarrow \infty$ and demands that the gravitational force should vanish there.
\end{itemize}

The above restrictions allow us to search for regular solutions to the $f(T)$-Yang-Mills system of equations. It is interesting to note that even when the conditions are written utilizing the equations of motion (for example, when using the equations of motion to rewrite the tetrad components in the Riemann-Christoffel and torsion tensor in terms of potentials) the conditions are all independent of $\alphaone$ and $\alphatwo$, and are therefore shared with TEGR and general relativity. Another related issue in common with general relativity is the fact that if one wishes the gravitational effects to vanish at infinity (and of course not be imaginary) then the vanishing of $ R^{\hat{\alpha}}_{\;\hat{\beta}\hat{\gamma}\hat{\delta}}$ and $T^{\hat{\alpha}}_{\;\hat{\beta}\hat{\gamma}}$ as $r\rightarrow \infty$ can be satisfied with the condition that $V'(r\rightarrow \infty)=0$ with some sufficiently high power. This may be seen from the corresponding conditions in table 1. One can conclude from this that if the gravitation is localized then one way to achieve this is with solutions which possess no global electric charge, as such a charge, $Q_{\tinyrmsub{E}}$, is defined by

\begin{equation}
Q_{\tinyrmsub{E}} \propto\oint_{r \rightarrow \infty}  \left|
\left|*F\, \right|\right| = \lim_{r\rightarrow \infty} r^{2} V'\,. \label{eq:globelec}
\end{equation}
The norm here is taken with respect to the Cartan-Killing form constructed of the structure coefficients of the $su(2)$ algebra.

Similarly, the vanishing (and reality) of gravitational effects at infinity also may be satisfied, from table 1, by the condition $W(r\rightarrow \infty)=\frac{1}{e}$, which in turn implies that there is no global magnetic charge, $Q_{\tinyrmsub{M}}$, via
\begin{equation}
Q_{\tinyrmsub{M}} \propto \oint_{r \rightarrow \infty} \left|\left|F\,\right|\right| = \lim_{r\rightarrow \infty} \left(1-e^{2}W^{2}\right)\,. \label{eq:globmag}
\end{equation}
Of course, one does not need to limit the solutions to those whose gravitational effects vanish as one approaches infinity, and admittedly the criteria used at $\infty$ in table 1 are sufficient, but not necessary conditions. However, we find that the successful magnetic solutions having this gravitational vanishing criterion at infinity turn out to indeed meet the condition of no global magnetic charge, as defined by (\ref{eq:globmag}). Electric solutions are studied separately below, and will turn out to possess a global electric charge dictated by (\ref{eq:globelec}).

Utilizing the restrictions discussed above, the following set of mutually compatible conditions are chosen in order to proceed with the asymptotically vanishing evolutions in the next section:
\begin{align}
A'(0)=&0, \; B(0)=1, \; B'(0)=0,\nonumber \\
V(0)=& 0, \; W(0)=\frac{1}{e}, \; W'(0)=0 \, .  \label{eq:rzeroconds}
\end{align}
The Yang-Mills coupling will also eventually be set as $e=1$.

\subsection{The solutions}\label{subsec:sols}
We will use as our free control parameter the quantity $W''$ at $r=0$. When solutions are analyzed which also possess $V$ the free parameter $V'$ will also be present. These are the quantities which one is allowed to set freely. In other words, all quantities required for the evolution must either be set to specific values at the starting point near $r=0$, as dictated by table 1, or if they cannot be fully fixed, must be written as functions of $W''$ and $V'$ only, again with the aid of table 1. In order to make clear the method used to generate the solutions we will present the scheme in some detail here. We set $V'(0)=0$ in the first set of evolutions, and the resulting equations then turn out to yield that $V=0$ throughout the evolution. Hence the first set of solutions concentrated on are purely magnetic.

Our problem involves two unknown metric profile functions, $A$ and $B$, and two unknown potentials, $W$ and $V$. We must therefore choose four independent differential equations. We found it convenient to use the ${}_t{}^t$ and ${}_\varphi{}^\varphi$-components of the equations of motion (\ref{eq:graveoms}), together with the two equations of motion for the Yang-Mills field (\ref{eq:ymexplicita},ii), as this particular choice allows for simple extraction of the highest order derivatives of the unknowns. The resulting system of four coupled ordinary nonlinear differential equations
is second order in $A$, $W$, and $V$, and first order in $B$.

In order to perform numerical integration of the equations, we first map the problem onto a compact domain by introducing
the dimensionless radial variable 
\begin{equation}
x := \frac{r}{r_0 + r} \in [0,1)\, ,
\end{equation}
where $r_{0}=1$ is set. At this stage all equations are to be rewritten in terms of $x$ instead of $r$. Just as the original equations are singular at $r=0$, the resulting equations are manifestly singular at $x=0$ (center), and also at $x=1$ (spatial infinity). This means that the initial conditions for numerical integration must be specified at some intermediate point. The usual strategy, which we adopt, is to start integrating from a point near the center, e.g.\ $x=x_0=0.001$. The initial conditions at that point can be derived using the power expansion of the unknown functions. The overall scheme is somewhat complicated and therefore summarized in point form as follows:

\begin{itemize}

\item The four differential equations are written in the form
      \begin{align} \label{eq:foureqns}
         &\Scale[0.90]{0 = f_i ( x, A, A', A'', B, B', W, W', W'', V, V', V'')\, ,} \nonumber \\
         &\qquad i = 1,\dots,4,
      \end{align}
      where the primes now denote differentiation with respect to $x$.

\item Assuming the unknown functions $A$, $B$, $W$, and $V$, are analytic,
      they are expanded in a Taylor series, and the resulting series are plugged into (\ref{eq:foureqns}).

\item The right hand sides of (\ref{eq:foureqns})
      are expanded into powers of $x$
      up to a sufficiently high order (see below).
      The coefficients in the resulting power expansion of $f_i$
      now involve the coefficients of the Taylor expansions
      of the unknown functions. That is, they consist of the functions at $x=0$.

\item As $f_i$, $i=1,\dots,4$, must vanish at all $x$,
      all coefficients in the power expansions
      of these equations are required to vanish individually. The resulting set of conditions, together with the conditions
      in (\ref{eq:rzeroconds}) allow us to determine the coefficients in the power expansions of the unknown functions up to the desired order
      (provided that the $f_i$ are expanded up to sufficiently high order).

\end{itemize}

The coefficients in the expansion of the unknown functions,
derived through the above procedure, allow us to approximate the solution to the equations as follows

\begin{subequations}
\romansubs
{\allowdisplaybreaks\begin{align}
A(x)  \simeq & \Scale[0.90]{1+\frac{1}{2} x^2 \left(\frac{4 \
b^2}{r_0^2}+E_r^2\right) + x^3 \left(\frac{4 b^2}{r_0^2}+E_r^2\right)} \nonumber \\
&\Scale[0.90]{-\frac{1}{40 r_0^8} \left[x^4 \left(4 b^2+r_0^2 E_r^2\right) \left(240
\alphaone  b^4+r_0^4 \left(15 \alphaone  E_r^4-44 b^2\right) \right.\right.} \nonumber \\ 
&\Scale[0.90]{\left.\left.+120 \alphaone  b^2 r_0^2 E_r^2+r_0^6 \left(8 b e-11 E_r^2-60\right)\right)\right]} , \label{eq:ainit} \\[0.2cm]
B(x) \simeq & \Scale[0.90]{1+\frac{1}{2} x^2 \left(\frac{4 
b^2}{r_0^2}+E_r^2\right)+ x^3 \left(\frac{4 b^2}{r_0^2}+E_r^2\right)} \nonumber \\
&\Scale[0.90]{-\frac{1}{40 r_0^8}\left[x^4 \left(4 b^2+r_0^2 E_r^2\right) \left(480 
\alphaone  b^4+r_0^4 \left(30 \alphaone  E_r^4-60 b^2\right)\right. \right.}\nonumber \\
&\Scale[0.90]{\left.\left.+240 \alphaone  b^2 r_0^2  E_r^2+r_0^6 \left(16 b e-15 E_r^2-60\right)\right)\right]} , \label{eq:binit}\\[0.2cm]
W(x) \simeq & \Scale[1.00]{\frac{1}{e} -b x^2 -2 b x^3}    \nonumber \\
&\Scale[1.00]{-\frac{x^4 \left(8 b^3+b r_0^2 \left(-3 b e+2 
E_r^2+30\right)+e r_0^4 E_r^2\right)}{10 r_0^2}} \nonumber \\
& \Scale[1.00]{-\frac{2 x^5 \left(8 b^3+b r_0^2 \left(-3 b e+2 E_r^2+10\right)+e r_0^4 
E_r^2\right)}{5 r_0^2}} , \label{eq:winit}\\[0.2cm]
V(x) \simeq  & \Scale[0.90]{{-x E_r-x^2 E_r}} \nonumber \\
&\Scale[1.00]{-\frac{x^3 E_r \left(8 b^2+r_0^2 \left(-2 b e+2 
E_r^2+5\right)\right)}{5 r_0^2}}\nonumber \\
&\Scale[1.00]{-\frac{x^4 E_r \left(24 b^2+r_0^2 \left(-6 b e+6 E_r^2+5\right)\right)}{5 r_0^2}}\, , \label{eq:vinit}
\end{align}}
\end{subequations}
where for the control parameters we have used the notation
\begin{equation}
\frac{\dd^2}{\dd r^2} W(r) \Big|_{r=0} =: - 2 b,
\end{equation}
and 
\begin{equation}
\frac{\dd}{\dd r} V(r) \Big|_{r=0} =: - E_r,
\end{equation}
which corresponds to the electric fields at the center.
The derived expressions (\ref{eq:ainit}-iv) are used to compute the initial values of
$A$, $A'$, $B$, $W$, $W'$, $V$, and $V'$, at the starting point $x=x_0$.

Typically, during the numerical integration of the initial value problem,
the potential $W$ diverges
and the integration fails to reach infinity ($x=1$). This means that a solution for those particular values does not exist. However, it is possible to find the set of values of parameters $\alphaone$, $\alphatwo$, $b$, and $E_r$, for which the integration reaches much closer to infinity than for the nearby values of the parameters. Also, around these values of the parameters
the divergence of $W$ reverses its direction (sign). This is a common occurrence in numerical evolutions, where if the initial parameter is slightly above the value for which a solution exists, the divergence in the evolution is in one direction, and if the initial parameter is slightly too small from the correct value, the divergence is in the opposite direction. Careful tracing of the direction of the divergence of $W$, and of the $x$-coordinate value beyond which the numerical integration is no longer possible (the length of the run, denoted as $x_{\tinyrmsub{max}}$), reveals the precise values of parameters for which the solution that satisfies boundary conditions at infinity may exist.

Some of our results are shown in figures \ref{fig:neg} and \ref{fig:pos} for various values of the dimensionless quadratic coupling $e^{2}\alphaone$. In the figures, the smaller the value of the graph along the vertical axis, the closer to infinity the integrator was able to reach to solve the equations subject to the initial conditions. The hollow dots on the graph indicate the value of $b=W'/2$ (horizontal axis) at which regular solutions exist to the $f(T)$-Yang-Mills system. It is important to note that the variable plotted on the vertical axis is a logarithm of $(1-x_{\tinyrmsub{max}})$, and hence at these low points the integration actually reached very closely to $x=1$. Each curve represents a different value of $e^{2}\alphaone$, including $e^{2}\alphaone=0$ (TEGR). The results are summarized here:

\begin{itemize}

\item Setting $\alphaone=\alphatwo=E_r=0$ we are dealing
with the TEGR equations and our procedure fully reproduces the results of Bartnik and McKinnon.
This means that one can obtain solutions with one or more nodes in $W$ and with increasing resolution one can in principle resolve infinitely many successful evolutions. A countable infinity of solutions
indexed by the number of nodes in $W$ is known to exist in this case \cite{ref:BM}-\cite{ref:infinite2}. These are the famous Bartnik-McKinnon solitons of Einstein Yang-Mills theory. The corresponding values of $b$ are $0.453615$, $0.651601$, $0.696915$, $0.704753$. However, at more than five nodes,
the reliability of the numerical procedure becomes doubtful and therefore we do not present them. Also, no solutions exist for $b>0.72$. These values are all in full agreement with general relativity \cite{ref:BM}.

\item With $e^{2}\alphaone < 0$ and $\alphatwo=E_r=0$, our results reveal
that  only a finite number of solutions exists (see Fig. \ref{fig:neg}). As well, for $e^{2}\alphaone < -4.4$ no solutions exist, regardless of the value of the tuning parameter $b$. 

\item Solutions with positive $\alphaone$, and $\alphatwo=0=E_r$ are illustrated in figure \ref{fig:pos}. The values of $b$ for which the solutions exist are becoming larger. Another interesting phenomenon is noticed here. For small positive $e^{2}\alphaone$ the maximum value of $b$ for which solutions exist initially grows very quickly, going from the TEGR value of $b_{\tinyrmsub{max}}\approx 0.72$, to approximately $b_{\tinyrmsub{max}}=21.5$ for $e^{2}\alphaone \approx 0.3$ (not shown). (Although the number of solutions  at this value of $\alphaone$ is finite. Seven solutions were found in this vicinity.) As $e^{2}\alphaone$ increases from approximately $0.3$, the maximum value of $b$ for which a solution exists drops off. For example, from figure \ref{fig:pos} it may be seen that for $e^{2}\alphaone =0.8$ one has a solution at the value $b_{\tinyrmsub{max}}\approx 7.5$, whereas for $e^{2}\alphaone=1.6$ the maximum solution occurs at $b_{\tinyrmsub{max}}\approx 4.5$.  Here again for $e^{2}\alphaone \neq 0$ only a finite number solutions exist.

\end{itemize}

\begin{figure}[!ht]
\begin{center}
\includegraphics[width=\columnwidth, clip]{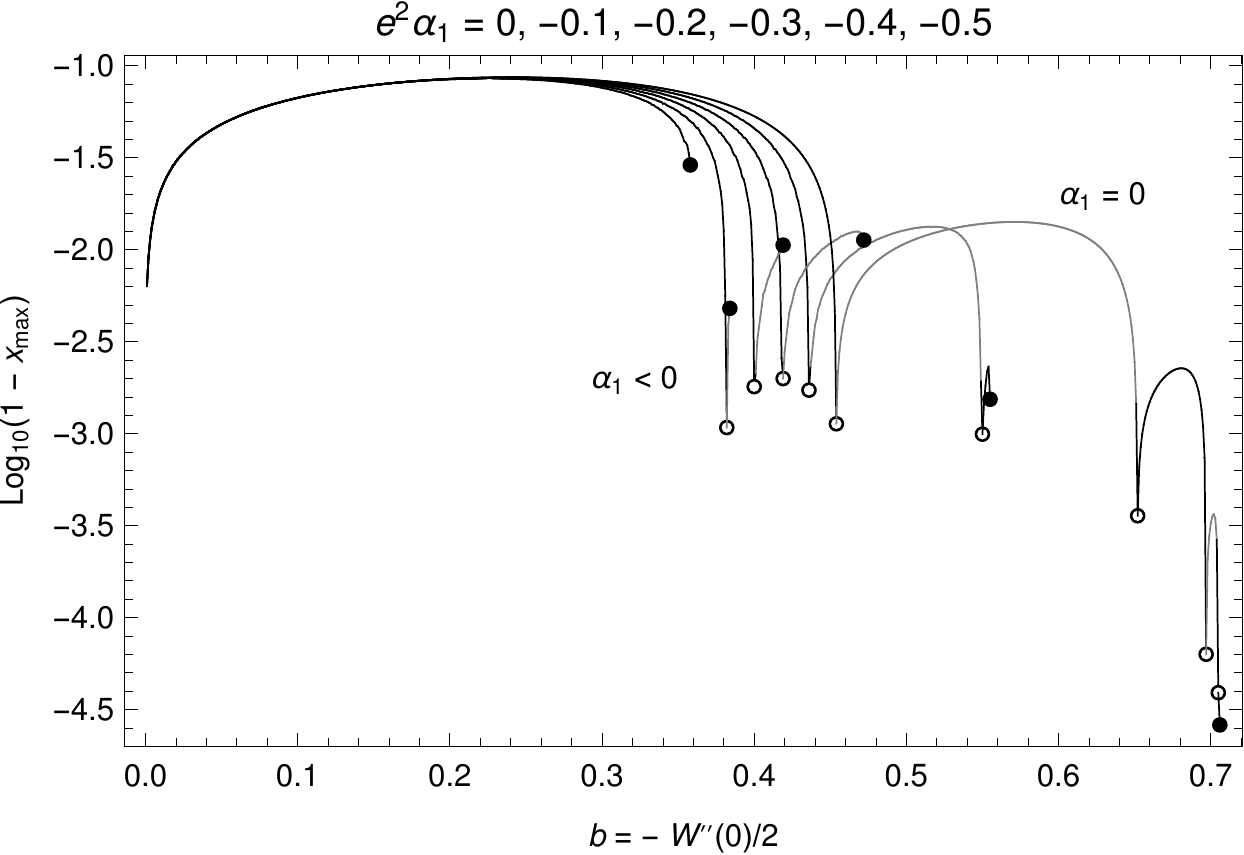}
\vspace{-0.75cm}\caption{\label{fig:neg}\small{Solutions with $e^{2}\alphaone=0$, $-0.1$, $-0.2$, $-0.3$, $-0.4$, $-0.5$,
$\alphatwo=0$. Here, in the horizontal axis label the derivatives of $W$ are with respect to $r$. The solid dots indicate the value of $b$ beyond which the integrator fails almost immediately. Regular solutions are indicated with hollow circles.
The black line indicates $W\to -\infty$ and the gray line indicates $W\to +\infty$ as $x\to x_{\max}$. See main text for explanation.}}
\end{center}
\end{figure}

\begin{figure}[!ht]
\begin{center}
\includegraphics[width=\columnwidth, clip]{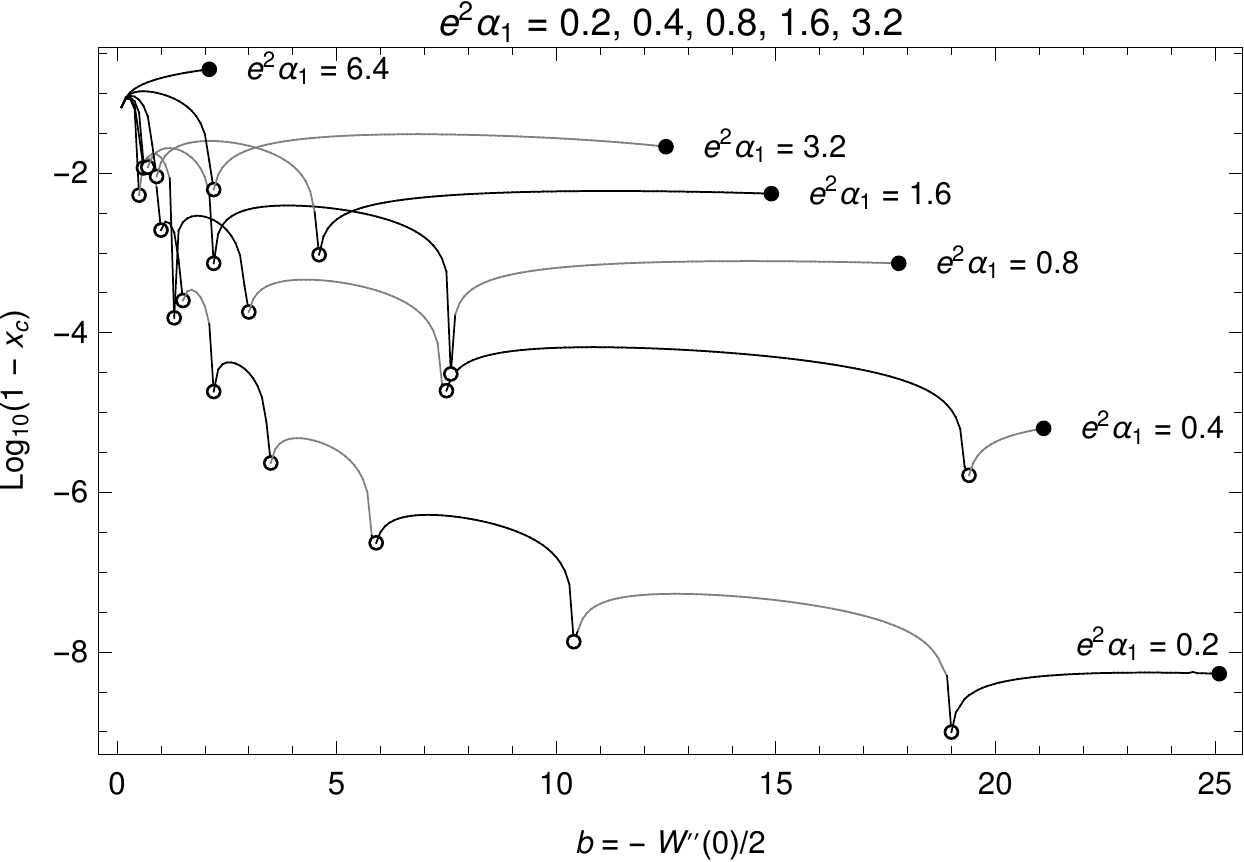}
\vspace{-0.75cm}\caption{\label{fig:pos}\small{Solutions with $e^{2}\alphaone=$ $0.2$, $0.4$, $0.8$, $1.6$, $3.2$, $6.4$, $\alphatwo=0$.}}
\end{center}
\end{figure}

Next we consider the scenarios where $\alphaone=0 =E_{r}$ and $\alphatwo \neq 0$, so that the cubic term in the action contributes. The equations here are rather complicated, and therefore the numerical solutions are time consuming to obtain, so a more limited number of $\alphatwo$ values are evolved. As well, the evolutions suffer from more numerical noise than their quadratic counter-parts. The results are illustrated in figure \ref{fig:beta}. We noted that for a magnitude of $e^{4}\alphatwo$ somewhere between $0.02$ and $0.025$ there exists a value beyond which there are no solutions if the magnitude of $e^{4}\alphatwo$ is increased. Therefore, similar to the case with $\alphaone$ there exists a cutoff value of $\alphatwo$ of the solutions as the nonlinear coupling increases in magnitude. It is interesting to note that this cutoff occurs for much smaller values than for the quadratic coupling, hinting that the existence of solutions is much more sensitive to higher power nonlinear torsion terms. We conjecture that there may exist some sufficiently high power of the torsion scalar in the Lagrangian such that no regular magnetic-only solutions will exist.

\begin{figure}[!ht]
\begin{center}
\includegraphics[width=\columnwidth, clip]{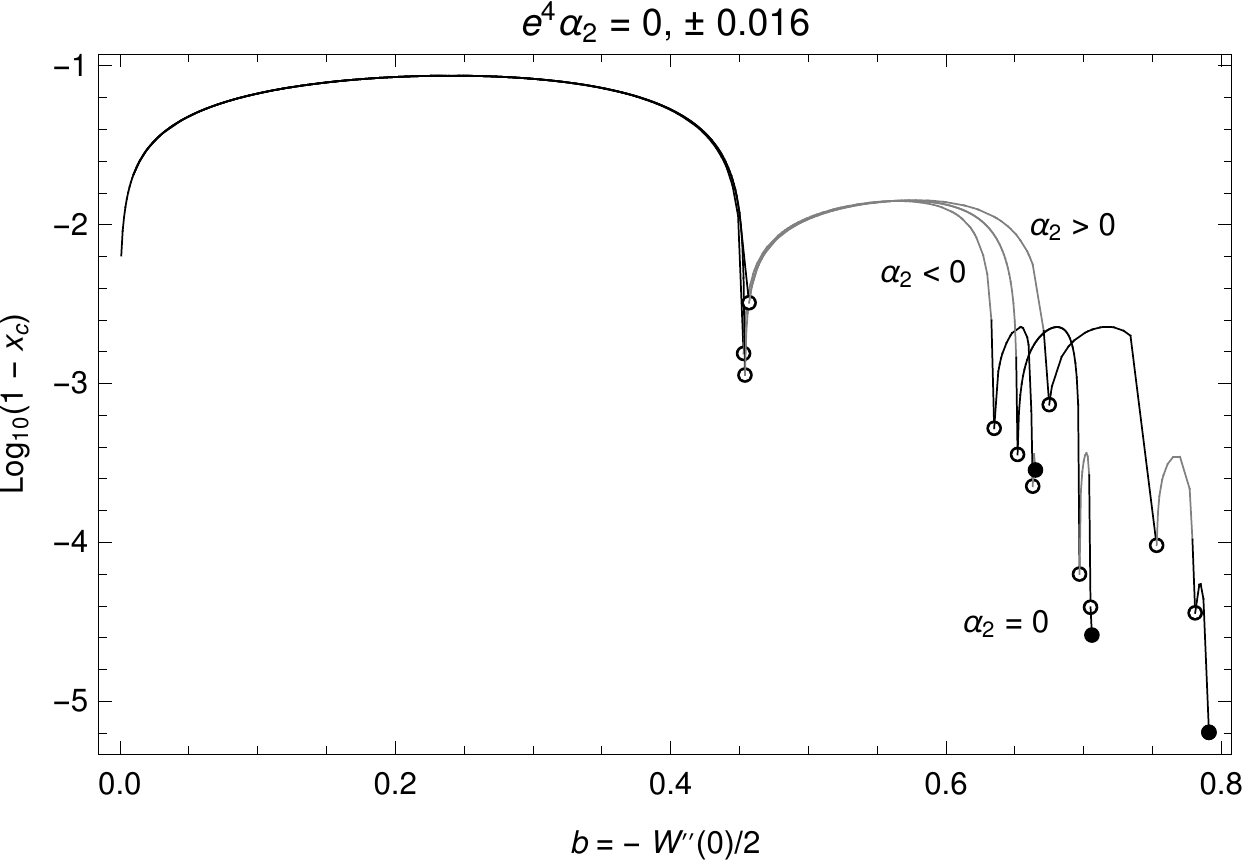}
\vspace{-0.75cm}\caption{\label{fig:beta}\small{Solutions with $e^{4}\alphatwo=$ $-0.016$, $0$ (TEGR), and $0.016$, $\alphaone=0$.}}
\end{center}
\end{figure}

Finally, for completion, solutions whose voltage, $V$, does not vanish everywhere are considered. For these, we must relax the previously chosen condition of $V'(0)=0$ and now this allows for Yang-Mills electric fields. One issue to note with non-zero $V'(0)$ is that one possesses a ``kink'' in the voltage at the origin. It turns out that this kink implies the presence of a point charge at the origin via the {gravitational} junction conditions of $f(T)$ gravity \cite{ref:jesse}, as the quantities demanded to be continuous in \cite{ref:jesse} are not all continuous at $r=0$ when $V'(0)\neq 0$. Mathematically this charge is manifest in (\ref{eq:globelec}), and figures \ref{fig:electegr} - \ref{fig:elecbetaneg} where the graph insets indicate that $r^{2}V'|_{r\to\infty}$ does not go to zero, and hence an electric charge is present according to (\ref{eq:globelec}). One other issue to note is that the effective mass of these solutions, as defined by the integral of $\rho$, the energy density of the Yang-Mills field, does not converge as strongly as in the purely magnetic counter-parts. This is an interesting issue of study in general relativity \cite{ref:eymmassstart},\cite{ref:mccormick}.

We note here that the family of regular solutions is no longer discrete, and that one may continuously deform the initial value of $E(0)_{r}$ to yield successful integrations, although if the value of $E(0)_{r}$ is too large in magnitude one runs into a regime with no solutions. There is therefore a critical value of $V'(0)$ separating a region of solutions and no solutions. What is surprising here is how insensitive to the value of $\alphaone$ the solutions are in the positive sector. It is found that for positive $\alphaone$ the evolution proceeds almost exactly as in TEGR even for very large values of positive $\alphaone$. In figures \ref{fig:electegr} and \ref{fig:elecalpha500} we show the solutions for the TEGR case as well as for $e^{2}\alphaone=500$ and note that the figures are almost identical. In fact, in order to lose the solution we found that one had to set $e^{2}\alphaone$ to a value of approximately $1050$. Any higher value than approximately this and there exists no solution. In the negative $\alphaone$ sector the critical value beyond which there are no solutions is only $-3.3$. A representative solution is shown in figure \ref{fig:elecalphaminus1}, where $e^{2}\alphaone=-1$.

The $\alphatwo$ dependence is again surprisingly sensitive to the existence of solutions. In this case, if $e^{4}\alphatwo$ is very small in magnitude, the solution is also very similar to the TEGR case, perhaps not unexpectedly. Representative solutions are shown in figures \ref{fig:elecbetapos} and \ref{fig:elecbetaneg}, for positive and negative $\alphatwo$ respectively. However even a very modest increase in the magnitude of $\alphatwo$ will destroy the solution. In the representative case the values of $\alphatwo$ which generated solutions restricts $e^{2}\alphatwo$ to the range $0.00026 > \alphatwo > -0.00026$. For all other values of initial conditions tested the range of $\alphatwo$ was also highly restricted to values in the close vicinity of the TEGR scenario. 

Finally, if $b$ at the origin is also not equal to zero we found that the qualitative behavior remains as in the plots \ref{fig:electegr}-\ref{fig:elecbetaneg} shown but the curvature of the oscillations in $W$ is softened.

\begin{figure}[H]
\begin{center}
\includegraphics[width=\columnwidth, clip]{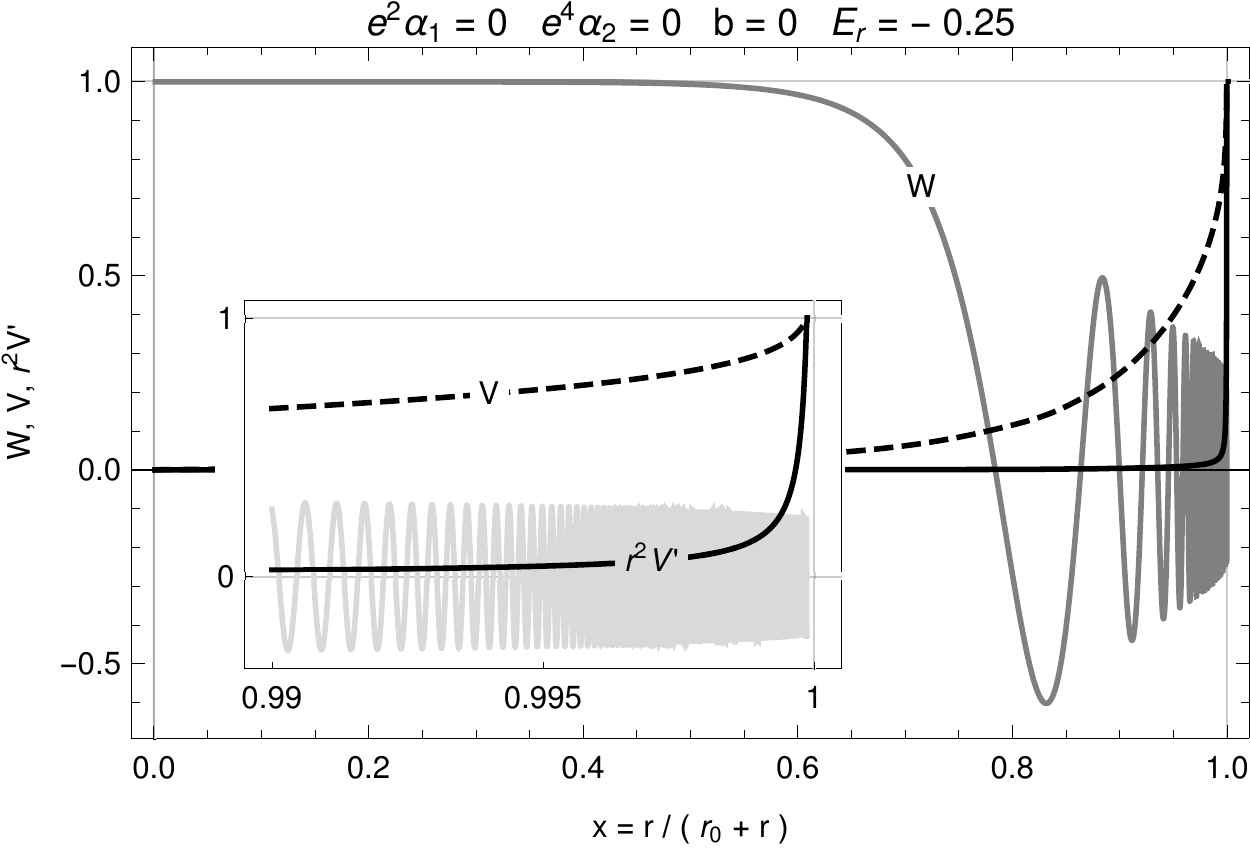}
\vspace{-0.75cm}\caption{\label{fig:electegr}\small{A representative TEGR scenario with electric field. For larger magnitude of $E_r$ the oscillations in $W$ become more damped, and for sufficiently high magnitude of $E_r$ there are no solutions.}}
\end{center}
\end{figure}

\vspace{-0.5cm}

\begin{figure}[H]
\begin{center}
\includegraphics[width=\columnwidth, clip]{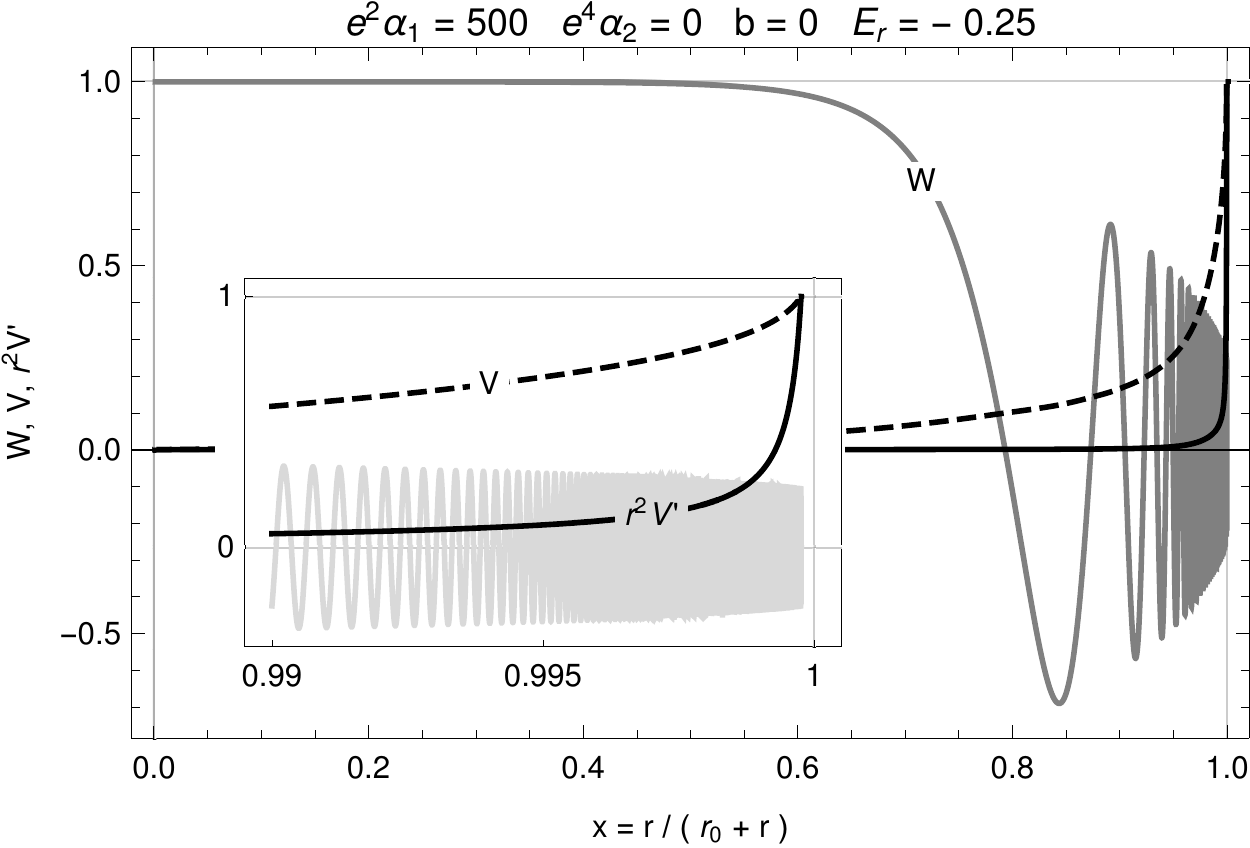}
\vspace{-0.75cm}\caption{\label{fig:elecalpha500}\small{A representative scenario with electric field with $e^{2}\alphaone=500$ and $e^{4}\alphatwo=0$. Note that this plot is essentially identical to the TEGR $(e^{2}\alphaone=0=e^{4}\alphatwo$) case.}}
\end{center}
\end{figure}

\vspace{-0.5cm} 

\begin{figure}[H]
\begin{center}
\includegraphics[width=\columnwidth, clip]{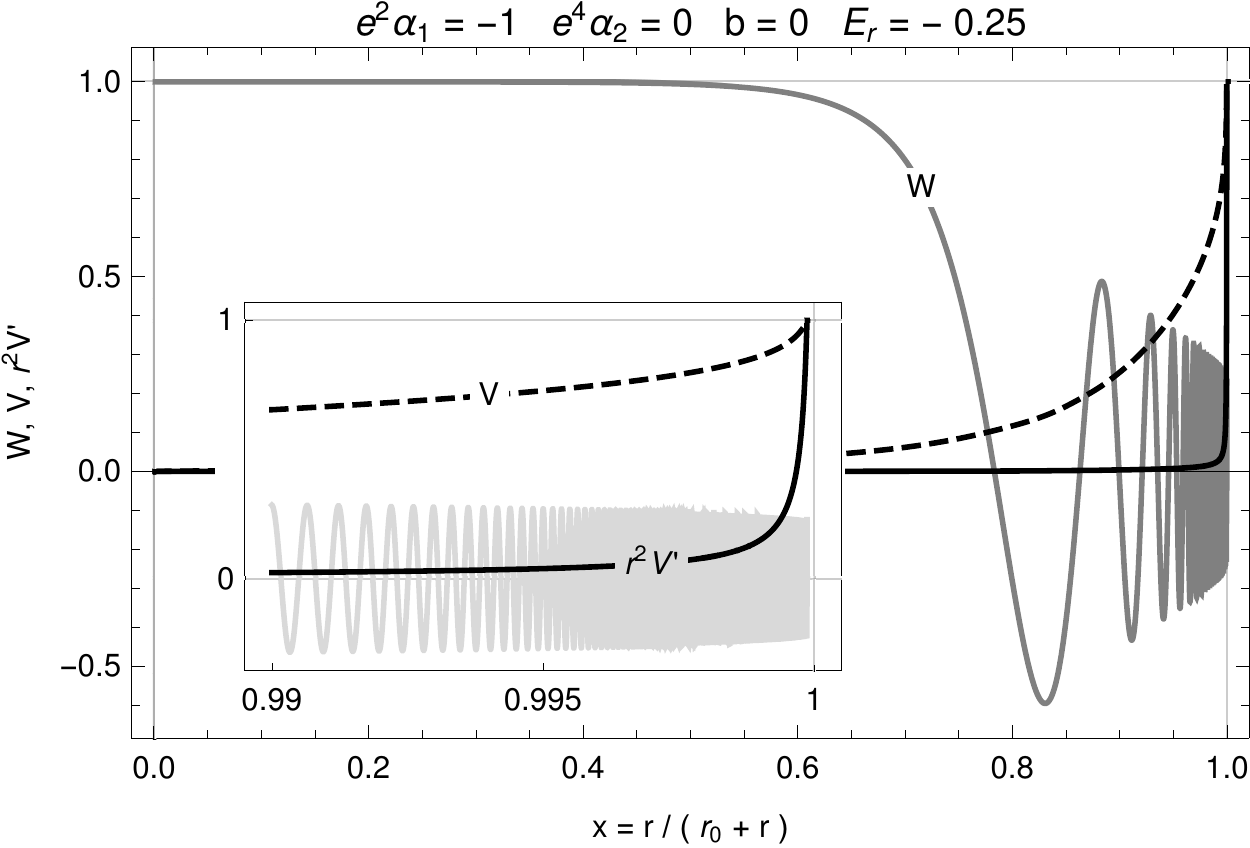}
\vspace{-0.75cm}\caption{\label{fig:elecalphaminus1}\small{A representative scenario with electric field with $e^{2}\alphaone=-1$ and $e^{4}\alphatwo=0$.}}
\end{center}
\end{figure}

\vspace{-0.5cm}
 
\begin{figure}[H]
\begin{center}
\includegraphics[width=\columnwidth, clip]{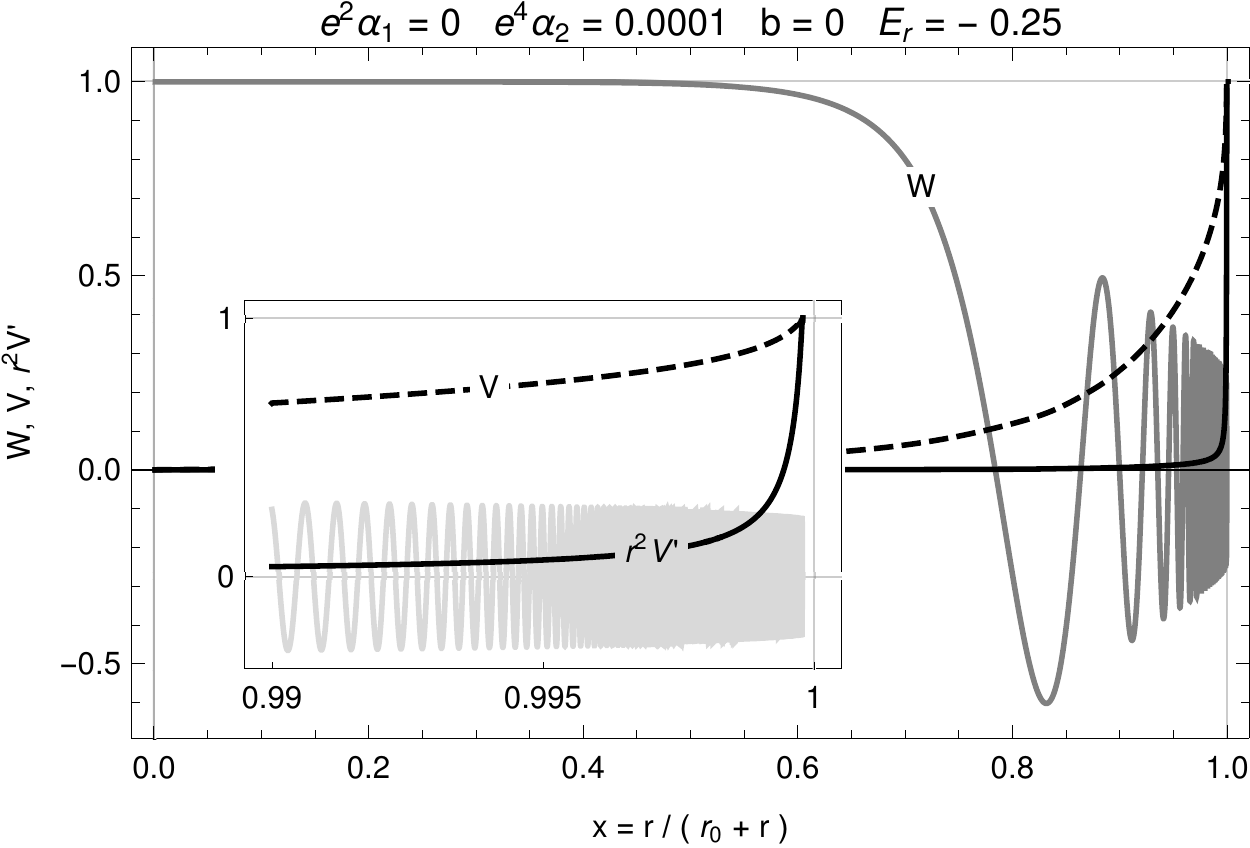}
\vspace{-0.75cm}\caption{\label{fig:elecbetapos}\small{A representative scenario with electric field with $e^{4}\alphatwo=0.0001$ and $e^{2}\alphaone=0$. If $e^{4}\alphatwo$ is larger than approximately this value, no solution exists.}}
\end{center}
\end{figure}

\vspace{-0.5cm}
 
\begin{figure}[H]
\begin{center}
\includegraphics[width=\columnwidth, clip]{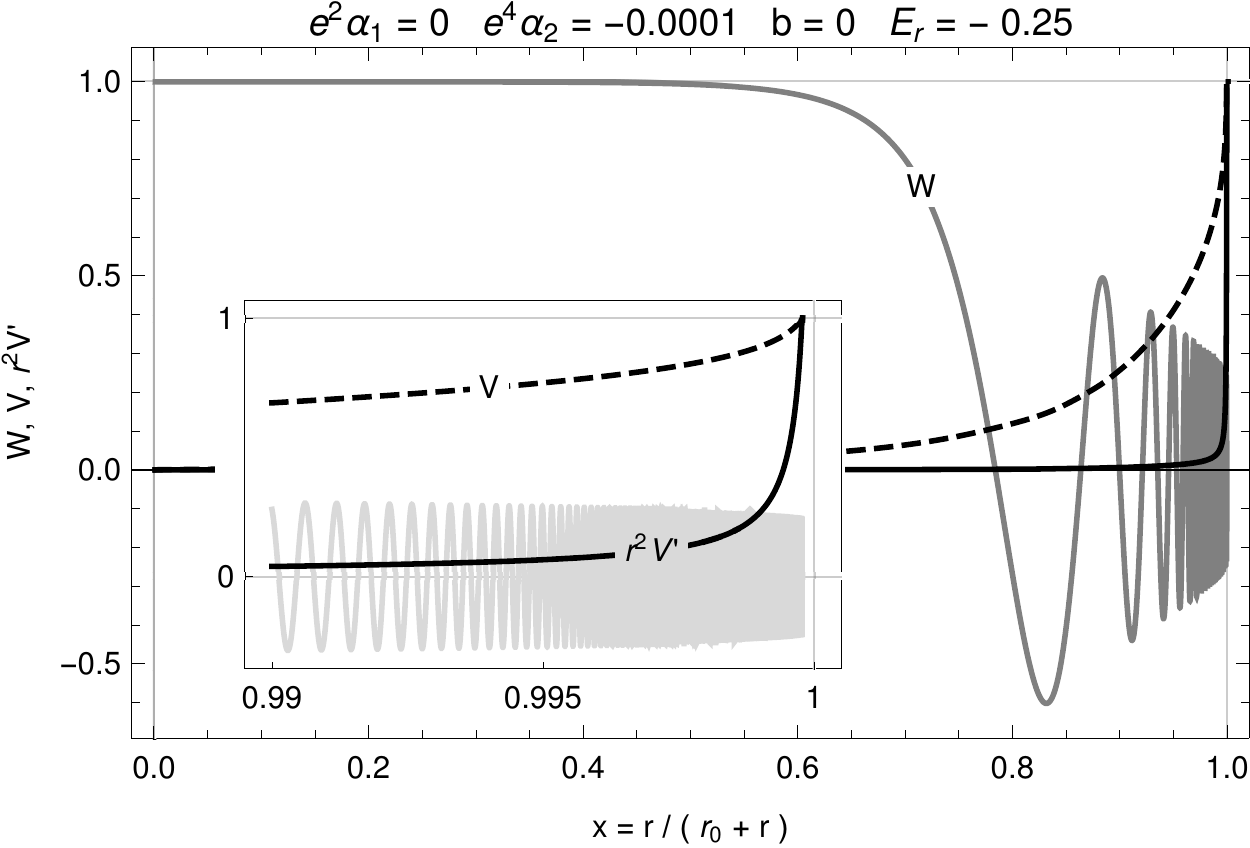}
\vspace{-0.75cm}\caption{\label{fig:elecbetaneg}\small{A representative scenario with electric field with $e^{4}\alphatwo=-0.0001$ and $e^{2}\alphaone=0$. This solution is very similar to $\alphatwo=0.0001$, but close inspection reveals that it is not exactly the same.}}
\end{center}
\end{figure}

\section{Concluding remarks}
In this manuscript the system of equations comprising of Lorentz covariant $f(T)$ gravity sourced by an $SU(2)$ Yang-Mills field was considered. Specifically for $f(T)$ we chose $f(T)=T+\alphaone/2\;T +\alphatwo/6\;T^{3}$, which could represent either an exact Lagrangian density, or else the low-order terms in a Lagrangian density analytic in $T$. In the purely magnetic sector we recover the Bartnik-McKinnon infinite family of soliton solutions for the case $\alphaone$ and $\alphatwo$ equal to zero, as expected. When the nonlinear coupling constants are not zero the number of solutions becomes finite. We find that there are critical values of $\alphaone$ and $\alphaone$ beyond which there are \emph{no} regular solutions. The domains where there exist regular solutions are given by $4.2 > e^{2} \alphaone > -4.4$ and $0.025 > e^{4} \alphatwo > -0.025$. It therefore turns out that the higher order torsion term in the action are actually rather more destructive to the presence of regular Yang-Mills solutions. 

Scenarios with electric fields were also studied, and were found to possess similar asymptotic mass issues as in general relativity. For the electric field scenarios the family of solutions becomes continuous. There is a value of the tuning parameter, $V'(0)$, above which there are no solutions. As well, there exist critical values of the coupling constants beyond which there are no solutions. In this case the restrictions for solutions were $1050 > e^{2} \alphaone > -3.3$ and $0.00026 > e^{4} \alphatwo > -0.00026$. As with the magnetic solutions, the sensitivity of existence to the cubic coupling is much stronger than with the quadratic coupling.

We also tested scenarios with both $\alphaone$ and $\alphatwo$ simultaneously not equal to zero and qualitatively similar results held. It may be possible therefore that in general, with a sufficiently high power of the torsion scalar present in the Lagrangian, one would find that there are no regular solutions to static $f(T)$-Yang-Mills theory.

\vspace{0.7cm}
\section*{Acknowledgments}
The authors are grateful to D.~Horvat, Z.~Naran\v{c}i\'{c} and M.~Sossich (U. Zagreb) for helpful discussions regarding this project, as well as to P.~Strossmayer. S.I. acknowledges the support of the VIF program of the University of Zagreb and the hospitality of SFU. AD is grateful for the kind hospitality of the physics department at FER (U. Zagreb) where much of this work was carried out. 

\PRLsep
\vspace{-0.080cm}

\linespread{0.6}
\bibliographystyle{unsrt}
 

\twocolumn[ 
\begin{@twocolumnfalse}
\appendix
\setcounter{equation}{0}
\renewcommand\theequation{A.\arabic{equation}}

\section*{Appendix - The gravitational equations of motion}
The terms in the equations of motion (\ref{eq:graveoms}) due to the presence of only $T$ in the action, in terms of our metric profile functions $A$ and $B$, are:
\begin{equation}
G^t{}_t = \frac{-B^3-2 r B'+B}{B^3 r^2} , \qquad
G^r{}_r = \frac{2 r A'-A B^2+A}{A B^2 r^2} , \qquad
G^\varphi{}_\varphi = \frac{B \left(r A''+A'\right)-B' \left(r A'+A\right)}{A B^3 r} .
\end{equation}
As we are using $f(T) = T + \frac{\alphaone}{2} T^2 + \frac{\alphatwo}{6} T^3 $, if we set $\alphatwo=0$, apart from the above terms
on left-hand-side of the $f(T)$ equations of motion we have the following terms proportional to $\alphaone$:
\begin{subequations}
\romansubs
{\allowdisplaybreaks\begin{align}
\tilde G^t{}_t =&  \frac{\alphaone}{A^2 B^5 r^4} (B-1) \left(A (B-1) \left(B \left(A \left(B^2-6 B+5\right)-8 \
r^2 A''\right)+12 A r B'\right)+8 A (B-3) r^2 A' B' \right. \nonumber \\
&\left.+4 (B-1) B r^2 \left(A'\right)^2\right) , \\[0.2cm]
\tilde G^r{}_r  = &\frac{\alphaone  (B-1) \left(A^2 (B+3) (B-1)^2-4 (B-3) r^2 \left(A'\right)^2-12 
A (B-1) r A'\right)}{A^2 B^4 r^4} , \\[0.2cm]
\tilde G^\varphi{}_\varphi  = & \frac{\alphaone}{A^3 B^5 r^4} \left(-A^2 (B-1)^2 \left(B \left(6 r^2 A''+A \left(B^2+2 \
B-3\right)\right)-6 A r B'\right)+4 A (3-2 B) r^3 \left(A'\right)^2 B' \right. \nonumber \\
&\left.-4
(B-1) B r^3 \left(A'\right)^3+2 A (B-1) r A' \left(B \left(4 r^2 A''+A \left(2 B^2-B-1\right)\right)+3 A (B-3) r B'\right)\right) ,
\end{align}}
\end{subequations}
while if we set $\alphaone=0$ and include $\alphatwo$, we have terms proportional to $\alphatwo$:
\begin{subequations}
\romansubs
{\allowdisplaybreaks\begin{align}
\tilde G^t{}_t = & -\frac{2 \alphatwo}{3 A^3 B^7 r^6}  (B-1)^2 \left(A (B-1)-2 r A'\right) \left(A (B-1) \left(B \
\left(A \left(B^2-20 B+19\right)-24 r^2 A''\right)+30 A r B'\right) \right. \nonumber \\
& \left. +2 A r A' \left(B \left(B^2+4 B-5\right)+6 (2 B-5) r B'\right)+16 (B-1) B r^2 \left(A'\right)^2\right) , \\[0.2cm]
\tilde G^r{}_r = & -\frac{2 \alphatwo  (B-1)^2 \left(A (B-1)-2 r A'\right)^2 \left(2 (2 B-5) r A'+A \
\left(B^2+4 B-5\right)\right)}{3 A^3 B^6 r^6} , \\[0.2cm]
\tilde G^\varphi{}_\varphi = & \frac{2 \alphatwo}{3 A^4 B^7 r^6} (B-1) \left(A (B-1)-2 r A'\right) \left(A^2 (B-1)^2 \left(B \
\left(15 r^2 A''+2 A \left(B^2+4 B-5\right)\right)-15 A r B'\right) \right. \nonumber \\
& -2 A r^2 
\left(A'\right)^2 \left(B \left(2 B^2-7 B+5\right)+3 (5-3 B) r B'\right)+12 \
(B-1) B r^3 \left(A'\right)^3 \nonumber \\
&\left. -A (B-1) r A' \left(B \left(18 r^2 A''+A \
\left(8 B^2+5 B-13\right)\right)+15 A (B-3) r B'\right)\right) .
\end{align}}
\end{subequations}
\PRLsep
\end{@twocolumnfalse}
]


} 
\end{document}